\documentclass[sigconf,balance=false]{acmart}
\usepackage{braket}
\usepackage{amsmath}
\usepackage{amsfonts}
\usepackage{enumerate}
\usepackage{algorithm}
\usepackage{algorithmic}
\usepackage{placeins}
\usepackage{makecell}

\renewcommand\footnotetextcopyrightpermission[1]{} 
\makeatletter
\def\@mkboth#1#2{} 
\let\markboth\@gobbletwo
\let\markright\@gobble
\AtBeginDocument{%
  \pagestyle{plain}
  \thispagestyle{plain}
}
\makeatother

\begin{document}
    \title{Data Obfuscation for Secure Use of Classical Values in Quantum Computation}

    \author{Amal Raj}
    \email{amal.raj@singaporetech.edu.sg}
    \orcid{0009-0001-4025-1045}
    \affiliation{%
      \institution{Singapore Institute of Technology}
      \country{Singapore}
    }
    
    \author{Vivek Balachandran}
    \email{vivek.b@singaporetech.edu.sg}
    \orcid{0000-0003-4847-7150}
    \affiliation{%
      \institution{Singapore Institute of Technology}
      \country{Singapore}
    }
    
    \renewcommand{\shortauthors}{Raj and Balachandran}

    \begin{abstract}
        Quantum computing often requires classical data to be supplied to execution environments that may not be fully trusted or isolated. While encryption protects data at rest and in transit, it provides limited protection once computation begins, when classical values are encoded into quantum registers. This paper explores data obfuscation for protecting classical values during quantum computation. To the best of our knowledge, we present the first explicit data obfuscation technique designed to protect classical values during quantum execution. We propose an obfuscation technique that encodes sensitive data into structured quantum representations across multiple registers, avoiding direct exposure while preserving computational usability. Reversible quantum operations and amplitude amplification allow selective recovery of valid encodings without revealing the underlying data. We evaluate the feasibility of the proposed method through simulation and analyze its resource requirements and practical limitations. Our results highlight data obfuscation as a complementary security primitive for quantum computing.
    \end{abstract}


    \maketitle 

    \section{Introduction}\label{sec:Introduction}
        Quantum computing is increasingly deployed across a variety of execution environments, including shared hardware, experimental platforms, simulators, and cloud-based services \cite{ibm,braket}. In these settings, classical input data must be supplied to the quantum computing stack and encoded into quantum registers as part of the computation. While conventional data protection mechanisms such as encryption are effective at securing data at rest and in transit, they provide limited protection once computation begins, when classical values are explicitly represented within quantum registers. At this stage, sensitive data may be exposed to inspection, partial measurement, or misuse by an adversarial or curious execution environment, for example through debug hooks or diagnostic measurements. This gap between existing cryptographic protections and the realities of quantum execution motivates the need for mechanisms that protect data \emph{during} quantum computation, rather than only before or after execution.

        Obfuscation offers a complementary approach to encryption by transforming data and code into representations that preserve functionality while concealing underlying values. In classical software security, obfuscation is widely used to protect sensitive logic and data against the Man-at-the-End (MATE) attacker model \cite{mate}, where adversaries have direct access to executable code and runtime state. Translating this idea to quantum computation is non-trivial, as quantum programs operate on quantum states rather than classical memory, and partial observation can irreversibly affect computation through measurement-induced collapse. Existing research on quantum obfuscation has primarily focused on protecting quantum circuits and algorithms deployed on untrusted toolchains or hardware, using techniques such as compiler-level transformations \cite{dummy-gates}, split compilation \cite{split-compilation}, and logic locking \cite{eloq,opaque}. While effective for safeguarding quantum intellectual property, these approaches do not directly address the problem of obfuscating classical data values used during quantum computation. Similarly, quantum data hiding techniques conceal information in bipartite or multipartite states using entanglement and locality constraints \cite{divincenzo}, and are analyzed primarily as state-discrimination and access-control primitives rather than as mechanisms for supporting general computation over the hidden data.
        
        This work addresses this gap by exploring data obfuscation as a mechanism for protecting classical values during quantum computation. Instead of representing sensitive data directly, we encode a classical value into a structured quantum representation distributed across multiple registers. The encoding is defined implicitly through arithmetic constraints enforced by reversible quantum circuits, and Grover’s amplitude amplification is employed to selectively recover valid encodings. This design avoids direct exposure of the underlying data to the execution environment beyond what is revealed by the computation’s input–output behavior, while still allowing correct computation to proceed. We implement the proposed obfuscation technique using standard quantum primitives and evaluate its feasibility through simulation across multiple problem sizes. Our analysis characterizes qubit requirements, circuit depth, and execution overhead, highlighting both the practicality of the approach for small instances and the challenges posed by current quantum hardware. Overall, these results position data obfuscation as a complementary security primitive for quantum computing, applicable across a range of execution environments.

        \subsection{Contributions of This Paper}
            This paper makes the following contributions: 

            \begin{itemize}
            \item \textbf{Problem formulation for quantum data obfuscation: } We identify and formalize the problem of protecting classical values during quantum computation, highlighting a gap between existing encryption-based protections and the realities of execution-time data exposure in quantum systems.
            
            \item \textbf{Quantum data obfuscation construction: } We propose a quantum data obfuscation technique that encodes classical values into structured quantum representations distributed across multiple registers, using reversible arithmetic circuits and amplitude amplification to enable computation without direct data exposure.
            
            \item \textbf{Implementation and feasibility analysis: } We implement the proposed approach using standard quantum primitives and evaluate its feasibility through simulation, analyzing qubit requirements, circuit depth, and execution overhead across multiple problem sizes.
            \end{itemize}

        \subsection{Related Works}\label{sec:literature}
            Research on quantum obfuscation has developed along multiple directions, motivated by the need to protect quantum programs, execution environments, and sensitive information as quantum computing systems become more widely deployed. Early foundational work by Alagic and Fefferman established fundamental impossibility results for perfect black-box quantum circuit obfuscation, showing that general quantum circuits cannot be fully obfuscated under broad adversarial models involving multiple queries and access to multiple obfuscations \cite{alagic-fefferman}. Building on these results, Bartusek et al. proposed candidate constructions for indistinguishability obfuscation of pseudo-deterministic quantum circuits, supported by formal security arguments under quantum hardness assumptions \cite{bartusek}. These works establish the theoretical limits and possibilities of quantum obfuscation, primarily from a cryptographic and complexity-theoretic perspective.
            
            Subsequent research has focused on practical techniques for protecting quantum circuit intellectual property against threats arising from untrusted compilers and execution environments. Suresh et al. introduced dummy gate insertion, where SWAP gates are deliberately injected to corrupt circuit functionality during compilation and later removed by the designer to restore the original outputs, thereby obfuscating the circuit structure \cite{dummy-gates}. Saki et al. proposed split compilation, which partitions a quantum circuit across multiple compilation stages to prevent any single compiler from accessing the full circuit \cite{split-compilation}. Das and Ghosh developed a randomized reversible gate-based obfuscation technique that embeds random reversible subcircuits before compilation and appends their inverses afterward, making reverse engineering during compilation harder while maintaining functional correctness \cite{das-ghosh}. Raj and Balachandran proposed a hybrid approach that intentionally corrupts quantum outputs using additional gates and relies on lightweight classical post-processing to restore correctness, achieving improved obfuscation strength with reduced quantum overhead \cite{raj-balachandran}.
            
            Logic locking has also been adapted to quantum circuits as a means of protecting quantum intellectual property. Liu, John, and Wang introduced the E-LoQ framework, which compresses multiple key bits into single qubits and distributes key-dependent gates throughout the circuit, providing resistance to structural and functional reverse engineering with modest qubit overhead \cite{eloq}. Similarly, Rehman, Langford, and Liu proposed OPAQUE, a phase-based obfuscation scheme that embeds secret keys into gate parameters to conceal circuit functionality and structure \cite{opaque}. More recently, Bartake et al. introduced ObfusQate, an automated quantum program obfuscation framework that integrates multiple obfuscation strategies and demonstrates resilience against code extraction attacks, including those using large language models \cite{obfusqate}. While effective for protecting circuit implementations and program logic, these approaches primarily focus on obfuscating quantum code rather than classical data values used during computation.
            

            Beyond circuit obfuscation, several works extend quantum security mechanisms to related domains such as authentication and secure execution. Huang and Tang constructed a quantum state obfuscation scheme for unitary quantum programs, built around a functional quantum authentication scheme that allows authorized parties to learn specific functions of authenticated quantum states with simulation-based security \cite{huang-tang}. Zhang et al. introduced an encrypted-state quantum compilation scheme for quantum cloud platforms that applies quantum homomorphic encryption and quantum circuit obfuscation to enable computation over protected quantum states while limiting structural and output information leakage to cloud service providers \cite{zhang}. These approaches address authentication, execution integrity, and platform security, but do not explicitly consider the obfuscation of classical input data values during quantum computation.
            
            A related but fundamentally distinct line of research is quantum data hiding, introduced by DiVincenzo, Leung, and Terhal \cite{divincenzo}. Quantum data hiding schemes use entanglement to conceal classical or quantum information such that it is indistinguishable under local operations and classical communication (LOCC) \cite{locc}, providing information-theoretic security guarantees. Subsequent works have extended these ideas to multipartite systems, qubit hiding, and more general probabilistic theories \cite{hiding-quantum-data, ultimate-data-hiding}. However, quantum data hiding schemes are explicitly designed to prevent computation over the hidden data unless global quantum operations are permitted, and therefore address a different security objective than computational obfuscation techniques that aim to conceal data while still enabling controlled use during computation.
            
            Overall, existing work on quantum obfuscation and related security mechanisms can be broadly categorized into quantum circuit and program protection, authentication and secure execution, and information-theoretic data hiding. Despite significant progress in these areas, none explicitly address the computational obfuscation of classical data values used during quantum computation. In particular, existing circuit obfuscation techniques focus on algorithmic structure, while data hiding schemes preclude computation over concealed values. To the best of our knowledge, the use of quantum computational primitives to obfuscate classical numerical relationships while preserving computational usability remains unexplored. This gap motivates our investigation into quantum data obfuscation as a complementary security primitive for quantum computing.
        
        \subsection{Overview of the Paper}
            The paper is organized as follows. Section \ref{sec:background} provides an overview of the various concepts needed to understand the paper. Section \ref{sec:methodology} explains the methodology of the technique. The technique is illustrated through a sample case study in Section \ref{sec:case-study}. Evaluation metrics are shown in \ref{sec:evaluation}. Section \ref{sec:limitations} discusses the limitations of the present work and outlines possible directions for future research, while Section \ref{sec:conclusion} concludes the paper.
    
    \section{Background}\label{sec:background}
        \begin{enumerate}
            \item \textit{Qubit}: 
            Qubits, or quantum bits, form the fundamental unit of quantum information, analogous to bits in classical computation. Unlike a classical bit, which can exist exclusively in one of two states, $0$ or $1$, a qubit may exist in a linear superposition of both computational basis states, denoted $\ket{0}$ and $\ket{1}$. The general state of a single qubit is expressed as:
            
            \begin{equation*}
                \ket{\psi} = \alpha\ket{0} + \beta\ket{1},
            \end{equation*}
            
            where $\alpha, \beta \in \mathbb{C}$ are complex probability amplitudes satisfying the normalization condition $|\alpha|^2 + |\beta|^2 = 1$. The probabilities of obtaining outcomes $\ket{0}$ or $\ket{1}$ upon measurement in the computational basis are given by $|\alpha|^2$ and $|\beta|^2$, respectively. This capacity to encode information in coherent superpositions underpins quantum parallelism, a key resource exploited in quantum algorithms. Moreover, the extension to multi-qubit systems enables \textit{entanglement}, a non-classical correlation without classical analog, which is central to quantum cryptography, communication, and obfuscation \cite{nielsen-chuang}.
        
            \item \textit{Quantum Gate}: Quantum gates are the fundamental building blocks of quantum computation, analogous to Boolean logic gates in the classical paradigm. Formally, a quantum gate corresponds to a unitary operator $U$ acting on one or more qubits, preserving normalization and ensuring reversibility (i.e., $U^\dagger U = I$). Single-qubit gates include the Pauli matrices ($X, Y, Z$), the Hadamard gate ($H$) which generates superpositions, and the phase gates (such as $S$ and $T$ gates). Multi-qubit gates capture entangling operations, the most notable being the controlled-NOT (CNOT) gate, which enables non-classical correlations between qubits. Since any unitary operator can be decomposed into a sequence of such elementary gates, quantum gates serve as a universal primitive for constructing arbitrary quantum algorithms \cite{nielsen-chuang}.
        
            \item \textit{Quantum Circuit}: 
            A quantum circuit is a structured composition of qubits and quantum gates, serving as a model for quantum computation~\cite{kitaev}. The circuit begins with an initialized quantum register (often in the $\ket{0}^{\otimes n}$ state), undergoes a sequence of unitary transformations through applied gates, and concludes with projective measurements yielding classical outcomes. Quantum circuits provide both an intuitive diagrammatic representation and a rigorous mathematical abstraction of quantum algorithms.
        \end{enumerate}

        \subsection{Quantum Adder Circuits}
            Quantum adders constitute a fundamental class of arithmetic circuits in quantum computing, designed to perform integer addition on quantum registers \cite{adder}. Similar to their classical counterparts, they form the basis for more complex arithmetic operations, such as multiplication, modular reduction, and number-theoretic transforms. These operations are central to a wide range of quantum algorithms, including Shor’s factoring algorithm \cite{shor}, quantum linear system solvers \cite{quantum-linear-solver}, and cryptographic primitives. 
            
            Formally, given two $n$-qubit registers $\ket{a}$ and $\ket{b}$ representing integers $a,b \in \{0,1,\dots,2^n-1\}$, a quantum adder implements the reversible mapping
            
            \begin{equation*}
                \ket{a}\ket{b} \mapsto \ket{a}\ket{a+b \bmod 2^n},
            \end{equation*}
            
            ensuring reversibility and unitarity. Depending on the design, some adders also employ ancillary qubits for carry propagation or modular overflow detection.
            
            \subsubsection{Ripple-Carry Adders}
                Ripple-carry adders are direct quantum analogues of their classical counterparts \cite{adder}. They operate by computing each carry bit sequentially, starting from the least significant bit. At each stage, a controlled operation is used to update the carry qubit, which is then propagated forward to influence the computation of higher-order bits. While this design leads to a linear depth in the number of qubits $n$, it is conceptually simple and requires only a modest number of elementary gates, making it suitable for near-term implementations.
                
                The general structure of a ripple-carry adder consists of two phases:
                
                \begin{enumerate}
                    \item \textit{Carry Generation and Propagation:} Ancillary qubits are employed to compute carry information based on the input bits. Multi-controlled Toffoli gates are often used for this purpose.
                    \item \textit{Sum Computation:} Once carries are established, the sum bits are updated using CNOT operations conditioned on the input and carry states.
                \end{enumerate}
                
                Ripple-carry designs have been extensively studied, with optimizations proposed to minimize gate count and ancilla overhead. Among these, the \textbf{Cuccaro adder} is one of the most space-efficient constructions \cite{cuccaro}.
                
            \subsubsection{Cuccaro Adder}
                The Cuccaro adder \cite{cuccaro} is a seminal ripple-carry design notable for its simplicity and minimal ancillary qubit usage. Unlike earlier designs, which required a linear number of ancillas, the Cuccaro adder achieves full addition using only a single ancillary qubit to propagate carries. Its construction relies primarily on CNOT and Toffoli gates, both of which are standard primitives in fault-tolerant quantum architectures. The adder consists of three main components:
                \begin{enumerate}
                    \item \textit{Majority Gate (MAJ):} A reversible gate that computes the majority function of three inputs $(a_i, b_i, c_i)$, where $a_i, b_i$ are input bits and $c_i$ is the carry. The output is used to update the carry for the next stage.
                    \item \textit{UnMajority-and-Sum Gate (UMA):} The inverse operation of MAJ, which uncomputes the carry information while simultaneously producing the correct sum bit.
                    \item \textit{Carry Chain:} A sequence of MAJ gates that propagates the carry from the least significant to the most significant qubit, followed by UMA gates to restore ancillary states and output the final sum.
                \end{enumerate}
                
                Formally, for two $n$-bit registers $\ket{a}$ and $\ket{b}$ and an ancilla $\ket{c_0}$ initialized to $\ket{0}$, the Cuccaro adder implements:
                \begin{equation*}
                    \ket{a}\ket{b}\ket{0} \mapsto \ket{a}\ket{a+b}\ket{c_n},
                \end{equation*}
                where $c_n$ is the final carry-out bit. Importantly, all ancillary states except the final carry are uncomputed, preserving reversibility and ensuring that no residual garbage states remain.
                
                The Cuccaro adder’s gate complexity is $2n-1$ Toffoli gates and $5n-3$ CNOT gates, with a circuit depth linear in $n$. Its efficiency and elegance have made it a canonical benchmark in the study of quantum arithmetic, frequently used as a building block in modular exponentiation and number-theoretic algorithms.
    
        \subsection{Grover's Algorithm}\label{sec:grover}
            Grover’s search algorithm \cite{grover} is one of the fundamental quantum algorithms that demonstrates a provable speedup over classical approaches. It addresses the unstructured search problem: given an unsorted database of size $N$ and an oracle function $f:\{0,1\}^n \rightarrow \{0,1\}$, the task is to find an input $x^\ast$ such that $f(x^\ast)=1$. Classically, solving this problem requires $O(N)$ queries in the worst case, whereas Grover’s algorithm finds a solution using only $O(\sqrt{N})$ queries, offering a quadratic improvement.
        
            \subsubsection{High-Level Idea}
                Grover’s algorithm operates by amplifying the amplitude of the “marked” states (those $x^\ast$ for which $f(x^\ast)=1$) through a sequence of unitary transformations known as \emph{Grover iterations}. The algorithm proceeds as follows:
                
                \begin{enumerate}
                    \item \textit{Initialization:} Prepare an $n$-qubit register in the uniform superposition state
                    
                    \begin{equation*}
                        \ket{\psi_0} = \frac{1}{\sqrt{N}} \sum_{x=0}^{N-1} \ket{x}.
                    \end{equation*}
                    
                    \item \textit{Oracle Query:} Apply the oracle $O_f$, which flips the phase of the marked state:
                    
                    \begin{equation*}
                        O_f \ket{x} = (-1)^{f(x)}\ket{x}.
                    \end{equation*}
                    
                    \item \textit{Diffusion Operator:} Apply the Grover diffusion operator $D$, which inverts amplitudes about their average, thereby increasing the relative amplitude of marked states.
                    
                    \item \textit{Iteration:} Repeat the oracle and diffusion steps approximately $\frac{\pi}{4}\sqrt{\frac{N}{M}}$ times, where $M$ is the number of marked states.
                    
                    \item \textit{Measurement:} Measure the register, yielding the marked element $x^\ast$ with high probability.
                \end{enumerate}
    
            \subsubsection{Mathematical Intuition}
                The algorithm can be visualized geometrically as a rotation in a two-dimensional subspace spanned by:
                
                \begin{equation}
                    \ket{\alpha} = \frac{1}{\sqrt{N-M}} \sum_{x:f(x)=0} \ket{x}
                \end{equation}
                
                \begin{equation}
                    \ket{\beta} = \frac{1}{\sqrt{M}} \sum_{x:f(x)=1} \ket{x},
                \end{equation}
                
                Each Grover iteration rotates the state vector by an angle $2\theta$, where $\sin^2\theta = \frac{M}{N}$. After $O\left(\sqrt{\frac{N}{M}}\right)$ iterations, the system is close to $\ket{\beta}$, ensuring that measurement yields a marked element with high probability.
            
    \section{Methodology}\label{sec:methodology}
        This section outlines the methodology of our proposed quantum data obfuscation algorithm. The central idea is to conceal classical information within a hybrid quantum–classical framework by distributing it across quantum registers such that it is not directly observable. In our construction, the classical data consists of a natural number $N$. Through the obfuscation process, $N$ is not stored directly but instead encoded across three $n$-qubit quantum registers $x, y,$ and $z$, ensuring that the circuit enforces the relation $x+y+z = N$. The transformation hides $N$ across the registers but still makes it possible to amplify and recover the valid sums using quantum search.

        If $x,y,z$ each occupy $n$ qubits, the maximum representable sum is $3 \cdot (2^n - 1)$. Therefore, $n$ must satisfy $N \leq 3 \cdot (2^n - 1)$. Once $n$ is fixed, the circuit requires $3n + 5$ qubits in total: $3n$ for the addends, up to two carry qubits, two ancillas for the adder circuit and one for Grover amplification. The overall procedure is divided into three stages:
            
            \begin{enumerate}
                \item \textbf{Sum Computation:} Constructing a reversible circuit for $s = x+y+z$,
                \item \textbf{Grover Amplification:} Selectively amplifying solutions where $s=N$,
                \item \textbf{Measurement and Decoding:} Extracting valid decompositions of $N$.
            \end{enumerate}

        \subsection{Sum Computation}
            The first part of the circuit computes $s = x+y+z$ by cascading two reversible adders. This adder is well-suited for oracle constructions because it uncomputes intermediate carries, leaving no residual garbage qubits. The computation is structured as follows:
            
                \begin{itemize}
                    \item \textbf{Adder 1:} Computes $s_1 = x+y$, with the result stored in $y$ and a carry qubit $\texttt{cout\_0}$ holding the overflow bit.
                    \item \textbf{Adder 2:} Computes $s = s_1 + z$, where $z$ is extended to $(n+1)$ bits using an ancilla. A second carry qubit $\texttt{cout\_1}$ stores the final overflow bit.
                \end{itemize}

            Thus, the total qubit requirement for the adder module is $3n+4$, and the sum register $s$ is prepared for use in the Grover oracle.

        \subsection{Grover Amplification}
            After sum computation, we apply Grover’s algorithm to amplify states where $s=N$. This involves two key components:

            \subsubsection{Oracle Construction}
                The oracle compares the sum register $s$ with the classical target $N$. If equality holds, a dedicated ancilla qubit initialized in the $\ket{-}$ state undergoes a phase flip:
                
                \[
                    O_f\ket{s}\ket{-} = 
                    \begin{cases}
                        -\ket{s}\ket{-}, & \text{if } s=N,\\
                        \;\;\;\ket{s}\ket{-}, & \text{otherwise.}
                    \end{cases}
                \]
                The equality check is implemented using multi-controlled NOT gates with selective bit flips to match the binary representation of $N$.

            \subsubsection{Diffusion Operator}
                The diffusion operator inverts amplitudes about the mean:
                \begin{enumerate}
                    \item Apply Hadamard gates on the input register,
                    \item Apply Pauli-X gates on the same qubits,
                    \item Apply a multi-controlled $Z$ gate, with the input registers being the controls and ancilla being the target,
                    \item Undo the $X$ and $H$ gates.
                \end{enumerate}
            
                This transformation increases the amplitude of the marked states where $s=N$.

            \subsubsection{Number of Iterations}
                Let $M$ be the number of valid solutions to $x+y+z=N$, and $T=2^{3n}$ the total number of possible triplets. The optimal number of Grover iterations $R$ is
                \begin{equation}
                    R = \operatorname{round} \left(\frac{\pi}{4}\sqrt{\frac{T}{M}}\right)
                \end{equation}
                
                The value of $M$ can be determined using the inclusion–exclusion principle for bounded integer compositions:
                
                \begin{equation}
                    M = \sum_{j=0}^{3} (-1)^j \binom{3}{j} \binom{N - j\cdot 2^n + 2}{2}, \qquad N - j\cdot 2^n \geq 0.
                \end{equation}

        \subsection{Measurement and Decoding}
            After $R$ Grover iterations, we measure the $3n$ qubits corresponding to $x,y,z$. The observed bitstrings are decoded into integers, and the most frequently observed triplets correspond to valid decompositions of $N$. These outputs represent obfuscated encodings of $N$ as structured triplets, suitable for downstream cryptographic or functional obfuscation tasks.

    \section{Case Study: Obfuscating N=19}\label{sec:case-study}
\begin{figure}[h]
            \centering
            \includegraphics[width=1\linewidth]{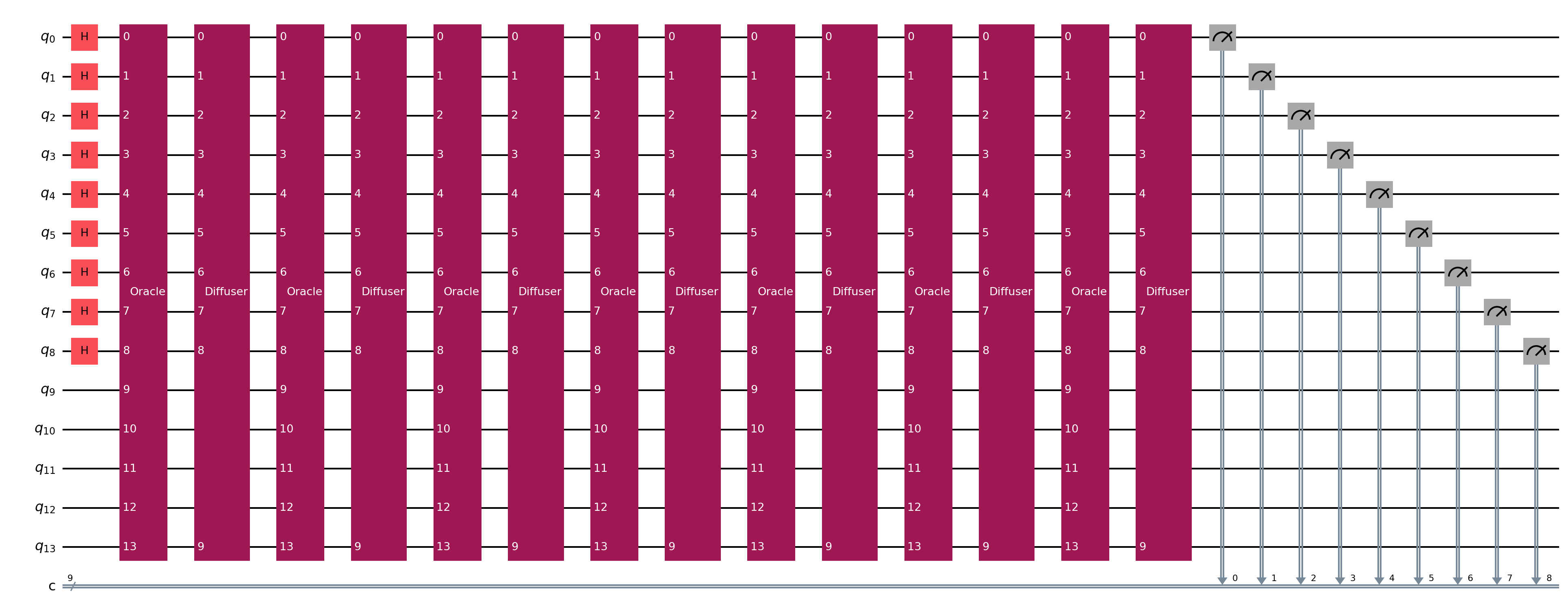}
            \caption{The final circuit for obfuscating $N=19$ as the sum of 3-bit numbers $x$, $y$, and $z$. The qubits $q_0$ to $q_8$ represent $x,y,z$, the first three being for $x$, next three for $y$ and last three for $z$ in little-endian format.}
            \label{fig:obfuscation-circuit}
        \end{figure}

        To illustrate the methodology, we present a concrete case study where the natural number $19$ is obfuscated as the sum of three integers $x,y,z$, each represented on $n=3$ qubits. This small-scale example demonstrates the practical construction of the quantum adder subcircuit and clarifies qubit allocation, ancilla reuse, and carry management. Fig. \ref{fig:obfuscation-circuit} shows a high-level overview of the circuit used.

        \subsection{Adder Architecture}
            The computation of $s = x+y+z$ is performed using two cascaded \textit{CDKMRippleCarryAdder} circuits from Qiskit \cite{cdkm}, each configured with \texttt{kind=`half'}. In this configuration:
            
            \begin{itemize}
                \item The \textbf{first adder} requires $2n$ qubits for the two operands, one ancilla qubit for internal carry propagation (returned to $\ket{0}$ after the operation), and one carry-out qubit to store the most significant bit (MSB).
                
                \item The \textbf{second adder} requires $n$ qubits for the third operand ($z$), reuses the $(n+1)$-qubit output of the first adder as its second input, and also reuses the ancilla of the first adder to extend $z$ to $(n+1)$ qubits. In addition, it introduces one new ancilla qubit for carry propagation and one carry-out qubit for the final MSB.
            \end{itemize}
        
            \subsubsection{Adder 1: Computing $x+y$}
                In Fig.~\ref{fig:adder}, the first adder (outlined in red) computes the partial sum $s_1 = x+y$:
                
                \begin{itemize}
                    \item Qubits $q_0,q_1,q_2$ represent the input $x$,
                    \item Qubits $q_3,q_4,q_5$ represent the input $y$,
                    \item The result $s_1$ overwrites register $y$,
                    \item The carry-out qubit $q_9$ stores the MSB of $s_1$,
                    \item Ancilla $q_{10}$ is used during computation but reset by the end.
                \end{itemize}
                
                Since $s_1$ may require up to $n+1=4$ qubits, the carry-out qubit $q_9$ is essential to store the overflow.
            
            \subsubsection{Adder 2: Computing $(x+y)+z$}
\begin{figure}[h]
                \centering
                \includegraphics[width=1\linewidth]{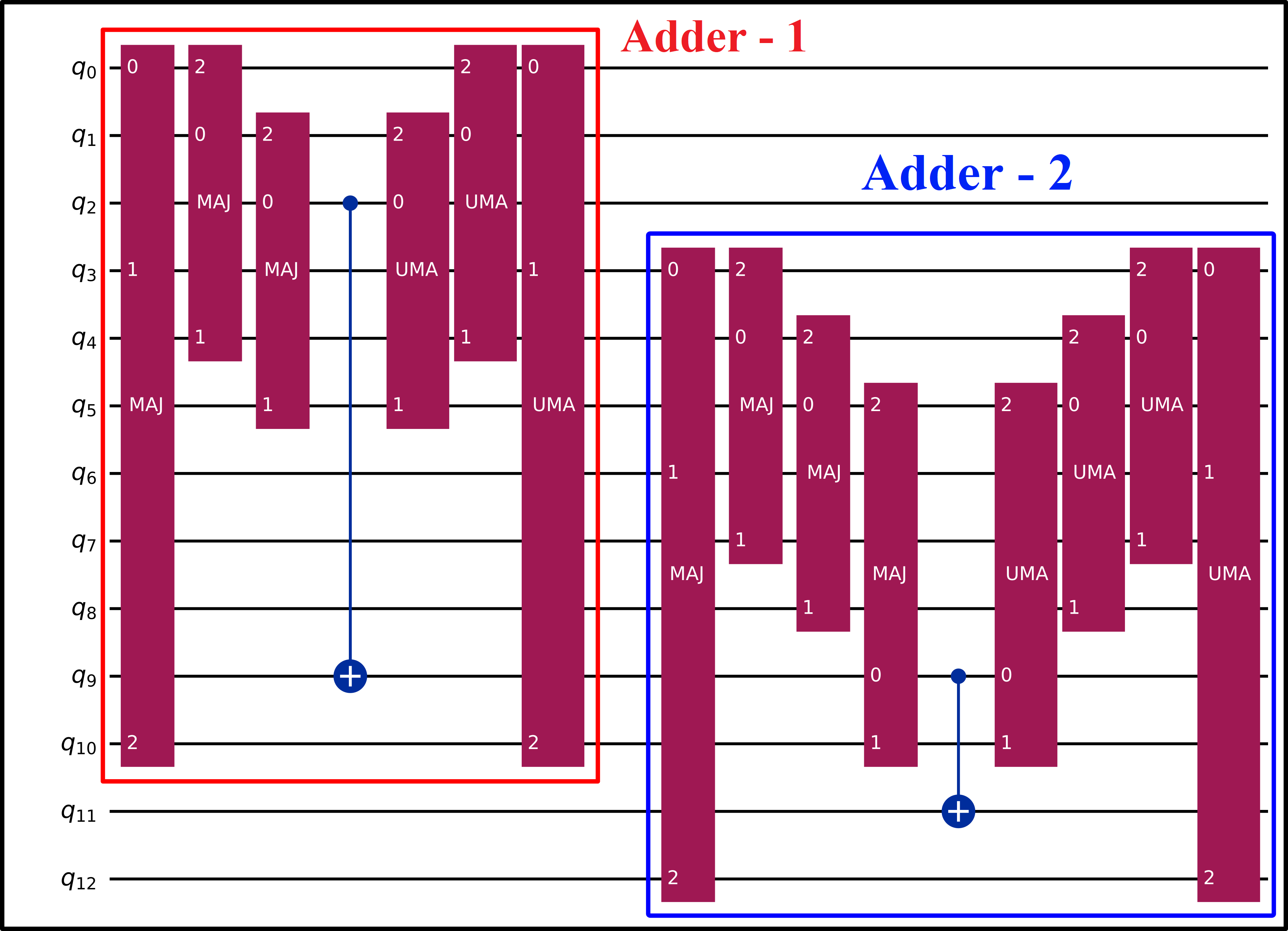}
                \caption{Quantum circuit for computing $s=x+y+z$ with $n=3$. Adder 1 (red) computes $s_1=x+y$, and Adder 2 (blue) computes $s=s_1+z$. Ancilla reuse is employed to extend $z$ to 4 bits in Adder 2.}
                \label{fig:adder}
            \end{figure}

                The second adder (outlined in blue in Fig.~\ref{fig:adder}) computes the final sum $s = s_1+z$:
                
                \begin{itemize}
                    \item $z$ is initially represented by qubits $q_6,q_7,q_8$.
                    \item To align with the $(n+1)$-bit width of $s_1$, $z$ is extended to 4 bits by prepending the previously used ancilla $q_{10}$.
                    \item Thus, the inputs to the second adder are:
                        \begin{itemize}
                            \item Addend 1: $(q_3,q_4,q_5,q_9)$ (the 4-bit sum $s_1$),
                            \item Addend 2: $(q_6,q_7,q_8,q_{10})$ (the 4-bit extended $z$),
                            \item Carry-out: $q_{11}$
                            \item Ancilla: $q_{12}$
                        \end{itemize}
                    \item The result $s = x+y+z$ is stored in $(q_6,q_7,q_8,q_{10},q_{11})$, where $q_{11}$ holds the final MSB.
                \end{itemize}
            
        \subsection{Grover Circuit}
            Building on the adder construction from Fig.~\ref{fig:adder}, we now extend the circuit with Grover’s algorithm to amplify those $(x,y,z)$ triplets that satisfy $x+y+z=19$, with each of $x,y,z$ encoded on 3 qubits. This case study demonstrates the oracle construction, diffuser design, and execution flow of the Grover stage.

        \subsubsection{Input Initialization}
            The input register consists of 9 qubits: $q_0,q_1,q_2$ for $x$, $q_3,q_4,q_5$ for $y$, and $q_6,q_7,q_8$ for $z$. All nine qubits are placed into equal superposition using Hadamard gates:
            \[
            \ket{\psi_0} = \frac{1}{\sqrt{512}} \sum_{x,y,z=0}^{7} \ket{x}\ket{y}\ket{z},
            \]
            where the search space is $2^{9}=512$ possible assignments. The Grover ancilla $q_{13}$ is initialized in the $\ket{-}$ state and serves as the phase flag.
            
        \subsubsection{Oracle Construction}
            The oracle is encapsulated as a gate to enable repeated application during Grover iterations. It has three components:
        
            \begin{enumerate}
                \item \textbf{Adder gate:} The previously constructed double half-CDKM adder is encapsulated as a single gate and appended. This maps $(x,y,z) \mapsto (x,y,z,s)$, where $s=x+y+z$ is exposed on the sum register $(q_6,q_7,q_8,q_{10},q_{11})$.
                
                \item \textbf{Query circuit:} To check whether $s=19$, we note that $19_{10} = 10011_{2}$ (5-bit representation). The query circuit applies $X$ gates to sum bits that are $0$ in this binary expansion, then performs a 5-controlled $X$ targeting the Grover ancilla $q_{13}$, and finally undoes the $X$ gates. This induces a phase flip only when the sum equals 19. This is shown in Fig. \ref{fig:query-circuit}.
                
                \begin{figure}[h]
                    \centering
                    \includegraphics[width=0.6\linewidth]{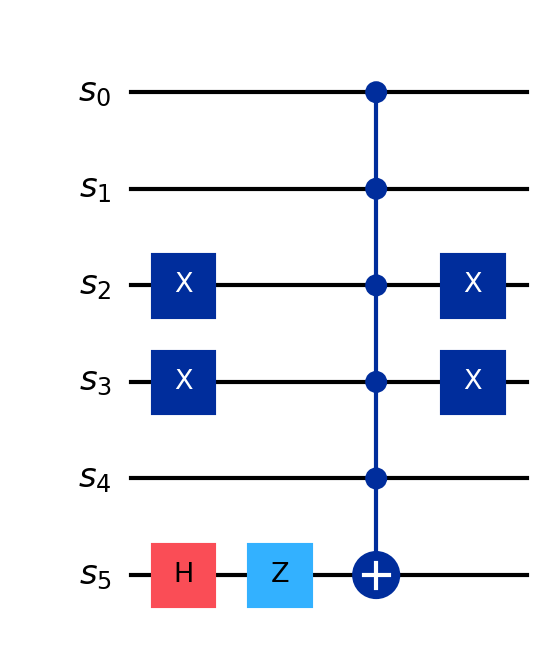}
                    \caption{Sub-circuit representing the query part of the oracle. Qubits $s_0$ to $s_4$ represent the sum, while $s_5$ is the ancilla qubit.}
                    \label{fig:query-circuit}
                \end{figure}
                
                \item \textbf{Uncomputation:} The inverse adder gate is appended to restore ancillas and remove garbage, leaving only the conditional phase on valid states.
            \end{enumerate}

\begin{figure}[h]
                \centering
                \includegraphics[width=0.5\linewidth]{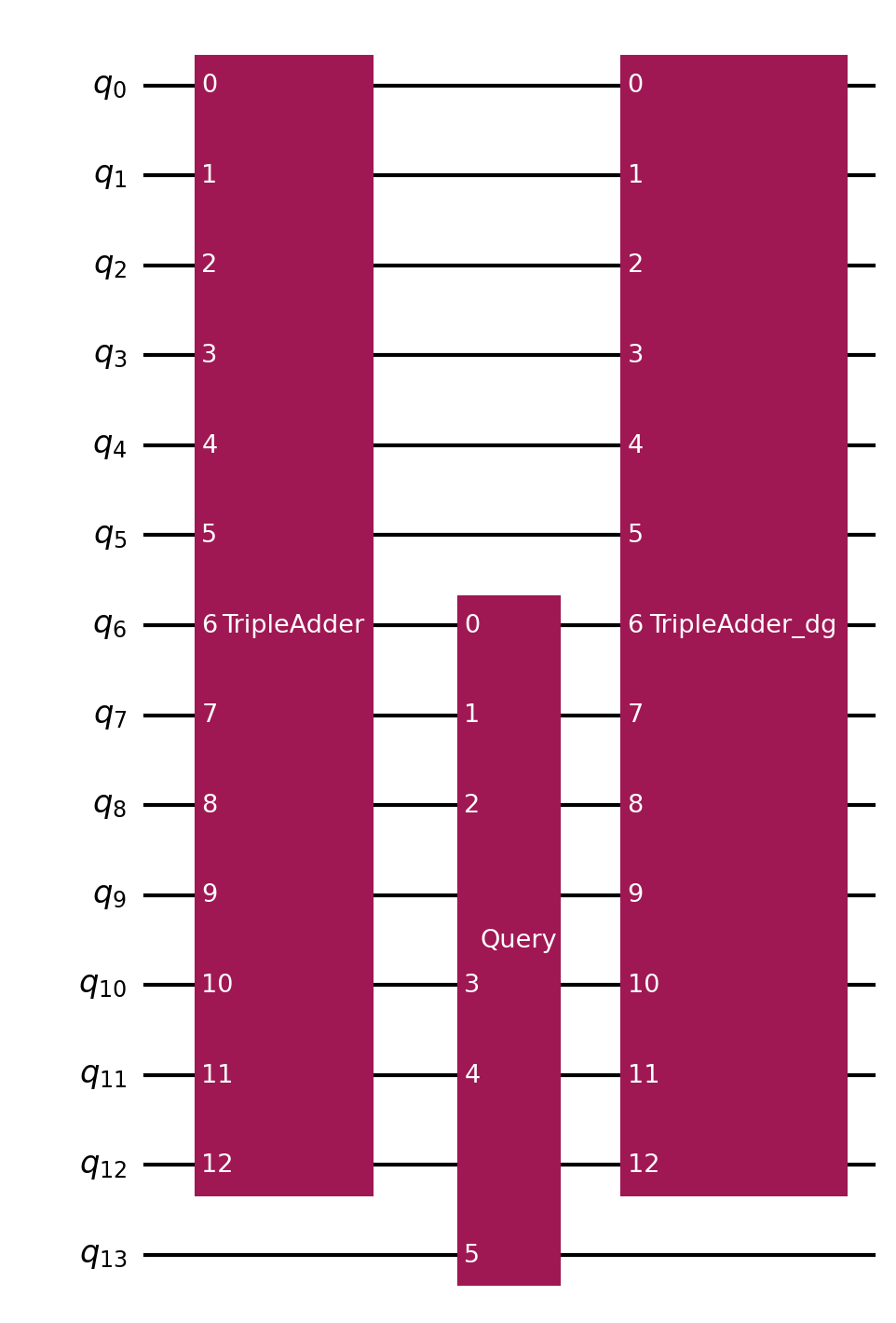}
                \caption{Oracle sub-circuit}
                \label{fig:oracle}
            \end{figure}

            The oracle circuit is shown in Fig. \ref{fig:oracle}.

            \FloatBarrier

        \subsubsection{Diffuser}
\begin{figure}[h]
                \centering
                \includegraphics[width=0.5\linewidth]{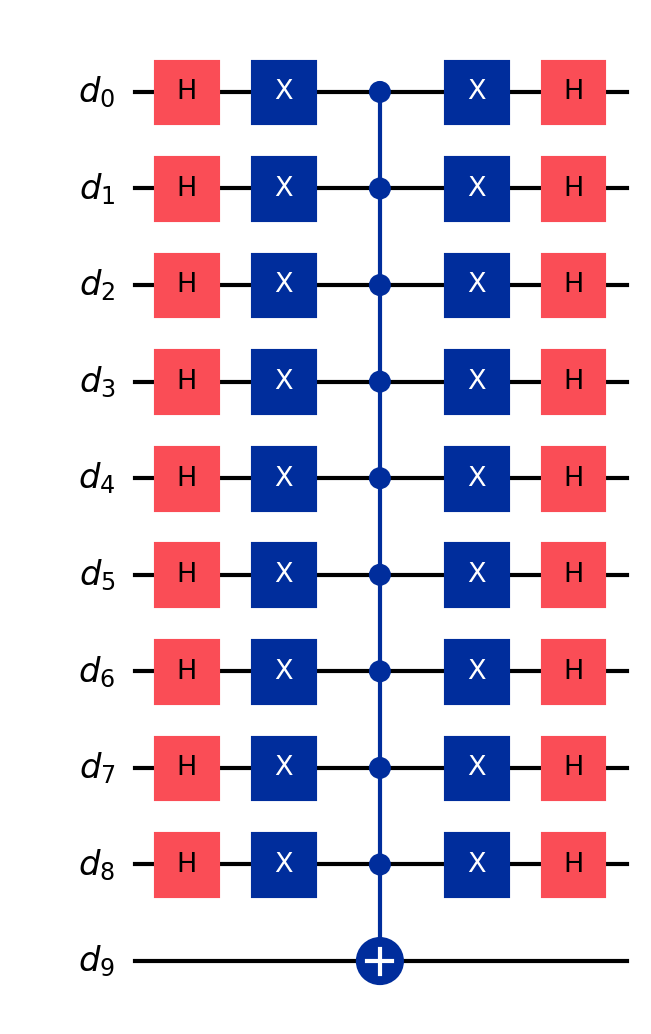}
                \caption{Diffuser sub-circuit. Qubits $d_0$ to $d_8$ represent the input registers, while $d_9$ is the ancilla.}
                \label{fig:diffuser}
            \end{figure}

            The diffuser is also implemented as a gate, acting only on the 9 input qubits and the Grover ancilla:
            \begin{enumerate}
                \item Apply Hadamards on $q_0$---$q_8$,
                \item Apply $X$ gates on $q_0$---$q_8$,
                \item Apply a multi-controlled $X$ with $q_0$---$q_8$ as controls and target $q_{13}$,
                \item Undo the $X$ and Hadamard gates.
            \end{enumerate}
            This realizes inversion about the mean for the 512-dimensional input space. The diffuser sub-circuit is shown in Fig. \ref{fig:diffuser}.

            \FloatBarrier

        \subsubsection{Grover Iterations for $N=19$}
            The number of valid solutions to $x+y+z=19$ with $0\leq x,y,z<8$ is $M=6$. With $T=512$ total states, the optimal number of Grover iterations is
            \[
            R \approx \frac{\pi}{4}\sqrt{\frac{T}{M}} = \frac{\pi}{4}\sqrt{\frac{512}{6}} \approx 7.26.
            \]
            We therefore apply $R=7$ iterations of the oracle and diffuser in sequence.

        \subsection{Measurement and Results}
            Finally, the 9 input qubits $q_0$–$q_8$ are measured. The measurement results are decoded into $(x,y,z)$ triples using the register mapping. The most frequent outcomes correspond to the valid decompositions of $19$. These constitute the obfuscated representations of $N$ as structured quantum states. An anonymized link for the code is available at \cite{github-repo}.

            \begin{figure}[h]
                \centering
                \includegraphics[width=1\linewidth]{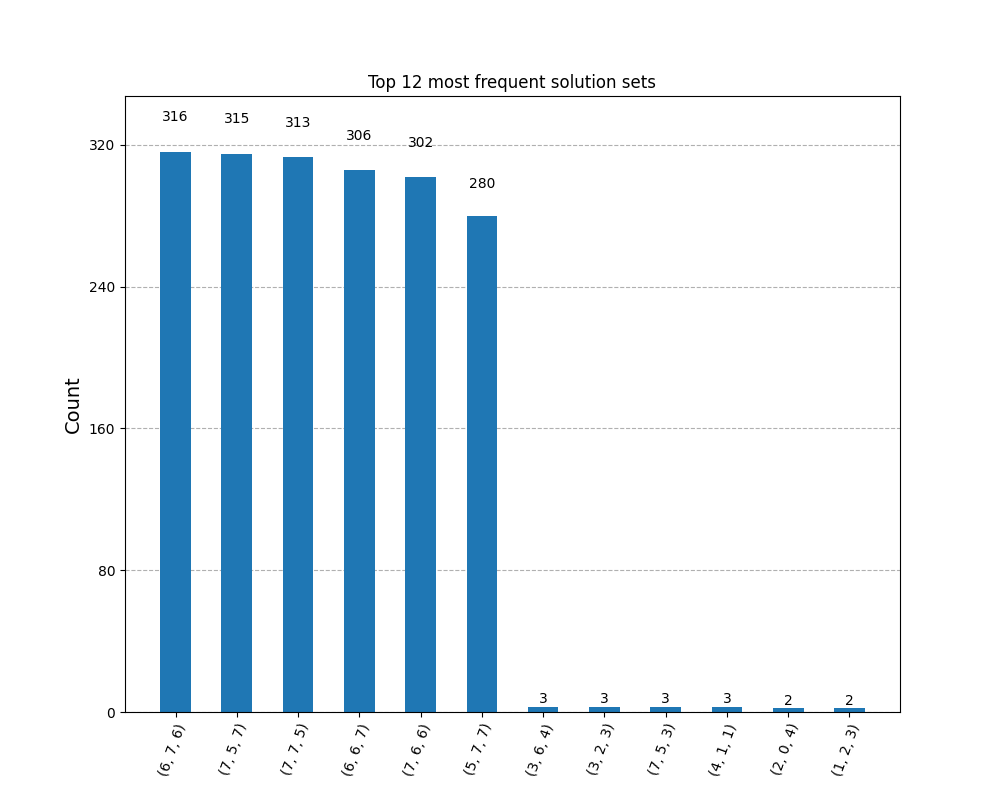}
                \caption{Histogram showing the frequency of top 12 combinations of $(x,y,z)$ decoded post-measurement. The triplets that sum to 19 occupy the highest counts (the six possible combinations occupying a total of 883 shots out of the total 1024), while others occur in negligible frequency.}
                \label{fig:histogram}
            \end{figure}
        \FloatBarrier
    
    \section{Evaluation}\label{sec:evaluation}
        \subsection{Implementation Setup}
            All simulations were conducted on a laptop equipped with an AMD Ryzen 5 5500U processor (2.10 GHz), Windows 11 Home OS, and 8 GB RAM. The software environment comprised Python 3.11.3 and Qiskit 2.0.2. Experiments were executed in a noise-free environment using the Qiskit AerSimulator to accurately measure circuit resource requirements. We selected benchmark values $N = 2^k-1$, for $k = 3, 4,5,...,8$, which correspond to natural boundaries in bit-width for the quantum decomposition registers $x,y,z$. This selection enabled a systematic study of circuit scalability from small $(N=7)$ to moderate ($N = 255$) problem sizes. Bit-width $n$
            was assigned such that all valid decompositions fit within n-bit registers, meeting the condition $N \leq 3 \cdot (2^n-1)$.
            
        \subsection{Evaluation Metrics}
            Table \ref{tab:benchmark-results} tabulates the register bit-width $n$, number of Grover iterations, number of qubits, circuit depth (after 3 levels of decomposition), run time,  total gate count (sum of all elementary gates used) and number of valid solutions (triplets $(x,y,z)$) for each target value $N$.
        
            \begin{table}[h!]
                \centering
                \caption{Benchmark results for varying $N$}
                \footnotesize
                \begin{tabular}{|c|c|c|c|c|c|c|c|}
                    \hline
                    N & \makecell{n\\(bits)} & \makecell{Grover\\Iterations} & Qubits & Depth & \makecell{Run\\Time (s)} & \makecell{Number\\of\\gates} & \makecell{Valid\\Solutions} \\
                    \hline
                    7 & 2 & 3 & 11 & 359 & 0.32 & 501 & 6 \\
                    \hline 
                    15 & 3 & 3 & 14 & 830 & 0.31 & 1029 & 28 \\
                    \hline 
                    31 & 4 & 5 & 17 & 2597 & 0.46 & 3017 & 120 \\
                    \hline
                    63 & 5 & 6 & 20 & 5168 & 2.23 & 5787 & 496 \\
                    \hline
                    127 & 6 & 9 & 23 & 11558 & 24.78 & 12645 & 2016 \\
                    \hline
                    255 & 7 & 13 & 26 & 23480 & 345.33 & 25293 & 8128 \\
                    \hline
                \end{tabular}
                \label{tab:benchmark-results}
            \end{table}
        
            It is evident that both depth and gate count increase rapidly with $N$, reflecting the cumulative cost of Grover iterations combined with cascaded ripple-carry adders in the oracle.

            The circuit depth shows a steep rise with $N$, starting from $359$ for $N=7$ and reaching $23480$ for $N=255$. This sharp growth highlights the combined complexity of the Cuccaro adder and Grover’s algorithm, as larger problem instances require more qubits and additional layers of sequential quantum operations. The number of gates follows a similar trajectory, further emphasizing the increasing resource overhead as $N$ scales.
            
            The runtime also grows with $N$, though the increase is less abrupt for small and moderate values. From $N=7$ to $N=63$, runtimes remain under $3$ seconds, while for $N=127$ and $N=255$, they rise sharply to $25$ seconds and $345$ seconds, respectively. Since runtime here refers to the execution time of the compiled circuit (excluding compilation time), these results illustrate that while classical simulation remains practical for small circuits, it becomes increasingly demanding as the size and depth of the quantum circuit grow.

            

        
        
    

        \subsection{Discussion}
            These results demonstrate that while the quantum obfuscation scheme is effective and scalable in principle, its resource requirements grow rapidly with $N$. For small values of $N$, both circuit depth and runtime are manageable, but for larger problems, the demands on quantum hardware and classical simulators become substantial. This underscores the importance of circuit optimization and motivates future work on more efficient quantum arithmetic and search techniques.

    \section{Limitations and Future Works}\label{sec:limitations}
        The principal limitation of our current approach lies in the depth and resource requirements of the constructed quantum circuits. As the size of the target integer $N$ increases, the circuit depth, gate count, and overall qubit requirements grow rapidly, resulting in circuits that are beyond the capability of current Noisy Intermediate-Scale Quantum (NISQ) \cite{preskill} hardware to execute reliably. While our simulations demonstrate conceptual feasibility, practical deployment will require significant advances in quantum hardware as well as further circuit optimizations. In addition, the present framework addresses only decomposition into exactly three summands, which, while illustrative and sufficient for proof of concept, may not represent the broadest range of obfuscation applications.

        Future work will explore alternative decomposition strategies for $N$, such as obfuscation schemes involving more than three summands or expressing $N$ as the solutions of Diophantine equations. Such extensions could allow for more complex and versatile obfuscation scenarios, enhancing applicability to a wider range of data types and cryptographic settings. Additional directions include optimizing quantum circuit design to reduce depth and gate count, investigating robustness with respect to realistic noise, and experimenting on real quantum hardware when available. The integration of cryptographic primitives and algorithm-specific obfuscation strategies also remains an open area for significantly strengthening the security and utility of this framework.

    \section{Conclusion}\label{sec:conclusion}
        We have presented a quantum-enabled data obfuscation scheme that encodes classical integers as quantum decomposition problems, leveraging Grover’s amplitude amplification to efficiently find and hide valid solutions. Our results demonstrate that, although significant resource barriers currently exist, this approach establishes a practical baseline for quantum data obfuscation and motivates further advances in both algorithms and hardware. The techniques developed herein extend the application of quantum obfuscation beyond software protection to the direct safeguarding of classical data, offering a novel cryptographic primitive for the emerging quantum era.
                    
    \bibliographystyle{ACM-Reference-Format}
    \bibliography{references}

\end{document}